\documentstyle[twoside,fleqn,espcrc2]{article}

\newcommand{\op}{{\bf D}}
\newcommand{\vect}[1]{{\bf #1}}
\newcommand{\calA}{{\cal A}}

\newcommand{\kernel}{{\calA^{[ 0\,j]}}}
\newcommand{\kernelmac}[1]{{\calA^{[ 0\, #1]}}}

\newcommand{\kernelone}{\kernelmac{1}}
\newcommand{\kernelarg}{{\calA^{[ 0\,j]}_{zx}}}
\newcommand{\kernelvec}{{\calA^{[ 0\,j]}_{\cdot,x}}}

\newcommand{\kernelmacvar}[2]{{\calA^{[ 0\,#1]}_{#2}}}

\newcommand{\cuberes}{|_{[x]}}

\def\tilde#1{\widetilde{#1}}
\def\xitilde{{\tilde{\vect{\xi}}}}

\def\rho{\varrho}
\def\epsilon{\varepsilon}
\def\phi{\varphi}
\def\be{\begin{equation}}
\def\ee{\end{equation}}

\newcommand{\AmS}{{\protect\the\textfont2
  A\kern-.1667em\lower.5ex\hbox{M}\kern-.125emS}}

\title{ISU --- Multigrid for computing propagators}

\author{M. B\"aker\address{
         II. Institut f\"ur Theoretische Physik
         der Universit\"at Hamburg,
         Luruper Chaussee 149, \\
         22761 Hamburg, Germany;
       e-mail: $<$baeker@x4u2.desy.de$>$}
        \thanks{Supported by Deutsche Forschungsgemeinschaft}
}
\begin{document}

\begin{abstract}
The Iteratively Smoothing Unigrid algorithm (ISU),
a new multigrid method for computing propagators in Lattice Gauge
Theory, is explained. The main idea is to compute good (i.e.\ smooth)
interpolation operators in an iterative way.
This method shows {\em no critical slowing down}
for the 2-dimensional Laplace equation in an SU(2) gauge field.
First results for the Dirac-operator are also shown.
\end{abstract}
\maketitle
\section{The problem}
The greatest obstacle for doing realistic (unquenched)
simulations of lattice QCD is
the large computer time required for the inversion of the
Dirac-operator due to critical slowing down. Multigrid algorithms,
which have been successfull for the solution of differential equations
describing ordered systems, have been studied for some time, but only with
limited success.

In \cite{previous} we proposed the ideas of a new multigrid method to
overcome these problems. This new algorithm, called {\em Iteratively smoothing
unigrid\/} or {\em ISU\/} has been described in detail in \cite{IJMPC}.
Here we want to review it shortly and on general grounds and want to
investigate its performance for the case of the 2-dimensional SU(2)
bosonic and fermionic propagator equation.

\section{The Unigrid}
In this section, the basic principles of the multigrid method are presented
from the unigrid point of view.

The two key observations are:

\begin{enumerate}
\item Standard relaxation algorithms (like Gau\ss-Seidel-relaxation)
applied on a lattice with lattice constant $a$ are
not efficient in reducing, but in smoothing the error on this scale.
(The error is defined as
the difference between the approximate and the exact solution.)
\item A function which is smooth on a length scale $a$ can be obtained by
interpolation from a grid with lattice constant $\propto a$.
\end{enumerate}

So the idea is to introduce, in addition to the fundamental lattice
$\Lambda^0$ on which the problem is defined, auxiliary layers $\Lambda^j$
with lattice constants $a_j = L_b^j a_0$, where $L_b$, the blocking factor, is
2 or 3. The last of these lattices, $\Lambda^N$, consists of only one point.
Points $x\in\Lambda^j$ can be identified with the corresponding points
in $\Lambda^0$. We define blocks $[x]$ in $\Lambda^0$ with sidelength
$2a_j-1$ and the point $x$ in the center.

Let the fundamental equation be
$\op\xi={\bf f}$.

To an approximate solution $\xitilde$ corresponds an error ${\bf e} =
\xi - \xitilde$.  We can cast the equation into the form $ \op\, {\bf
  e} = {\bf r}$, where ${\bf r} = {\bf f} - \op\,\xitilde$ is the
residual.

After relaxing on $\Lambda^0$, $\bf e$ is smooth on scale $a_0$. It
can therefore be obtained by interpolating it from a function ${\bf
  e}^1$ on $\Lambda^1$ with the help of the interpolation operator
$\kernelone$. $\kernel$ maps functions on $\Lambda^j$ to functions on
$\Lambda^0$, its adjoint $\kernel^\ast$ blocks functions from
$\Lambda^0$ to $\Lambda^j$. It has the property $\kernelarg=0$ for
$z\notin [x]$.  So we get ${\bf e}\approx \kernelone{\bf e}^1$,
resulting in the coarse grid equation
\begin{equation}\kernelone^\ast\, \op\,\kernelone\,{\bf e} =
  \kernelone^\ast\, {\bf r}\, \Longleftrightarrow\, \op^1\, {\bf e}^1 = {\bf
    r}^1\end{equation}. On this equation we relax again, thereby
  smoothing the error on scale $a_1$. We correct the approximation on
  the fundamental lattice $\xitilde \mapsto \xitilde - \kernelone\, {\bf
    e}^1$ and then we go to the next-coarser layer.  (This is the
  difference to a true multigrid method: There one goes from
  $\Lambda^1$ directly to the next layer, without correcting first on
  the fundamental lattice. This is advantageous, because the work
  involved is less, but that the ISU-algorithm cannot work as a true
  multigrid, see~\cite{IJMPC}.)

\section{The need for smooth operators}
The method described above can only work well if we know what is
meant by the smoothness of the error after relaxation. It can be
easily seen that in the context of lattice gauge theories, smoothness
is not a priori defined, because no unique way to compare function
values at different sites can be defined. Therefore we have to look for
a new, appropriate definition. The following two definitions are in most
cases equivalent.

\begin{description}
\item[Def.\ 1:]
A mode $\xi$ is smooth  on scale $\lambda$ iff
$\|{\bf L} \xi\| \ll \xi$,
$\bf L$ in units $\lambda=1$
\item[Def.\ 2:]
A mode is smooth on scale $\lambda$ iff
it is not efficiently reduced by relaxation on this scale.
\end{description}

That relaxation produces a smooth error not only according to the second,
but also to the first definition, can be seen from the fact that usually the
low-lying eigenmodes of $\op$ are the bad-converging ones. (See
\cite{Sokal} for a discussion of this and some caveats.)

To interpolate an error smooth in this sense, the interpolation operators
$\kernel$ have to be smooth themselves. So we have to find out how to
calculate smooth interpolation operators.
As the smoothest mode (according to def.\ 1) is the lowest eigenmode of
$\op$, it seems that we have to solve an equation as difficult as the one we
started with to find the $\kernel$.

\section{The iteratively smoothing unigrid}
But this is not true because of the following observation:

It is easier to calculate the shape of a mode which converges badly in a given
iteration scheme than to reduce it directly with this scheme.

This can be seen from the following example: Imagine that $\op$ only has
one bad-converging mode $\psi$. Then we can compute this easily by trying
to solve $\op\xi=0$, because in this case the approximate solution
equals the error, which converges fast to a (small) multiple of $\psi$.

The second problem solving strategy is to solve our problem by reducing
it to many similar problems which can be solved step by step.

The ISU algorithm for calculating good interpolation operators $\kernel$
works as follows:

To calculate $\kernelvec$ on a support$[x]$:

Solve $\op\cuberes \kernelarg = \epsilon_0 \kernelmacvar{j}{z^\prime,x}$
via inverse iteration, using all interpolation operators on layers
$\Lambda^k$ with $k<j$. (The righthandside could also be taken to be zero
as in the above example, according to def.\ 2 of smoothness.)
$\cuberes$ here means restriction to the block $[x]$, using e.g.~Dirichlet
boundary conditions.

The crucial point here is that the already calculated operators on the
finer layers are used for the iteration, otherwise there would be
many bad-converging modes and the calculation of $\kernel$ would be slow.

\section{Performance of ISU}
We studied the performance of this algorithm in a two-dimensional
SU(2) lattice gauge field, using the equation
\begin{equation} (\op -\epsilon_0 +\delta m^2) \xi = {\bf f}
  \quad ,\end{equation} where $\op$ was chosen to be the negative
  Laplace (bosonic case) or the negative squared staggered
  Dirac operator. $\epsilon_0$ is the lowest eigenvalue of the
  operator which was subtracted to be able to tune criticality by
  changing the parameter $\delta m^2$. This subtraction is necessary
  for the Laplace operator (because its lowest eigenvalue can be
  large) and it eases the analysis of the critical behaviour also for
  the Dirac case.

  We measured the asymptotic convergence rate $\tau$, i.e.~the number
  of iterations needed to reduce the error by a factor $e$, where by
  an iteration we mean visiting each layer of the unigrid twice in a
  so-called V(1,1)-cycle (see \cite{Brandt}).

In the bosonic case we studied the convergence rate on grids of sizes
$32^2$--$128^2$ at various values of the inverse coupling $\beta$.
It was found, that $\tau$ was approximately 1, independent of the
lattice size, $\beta$, and $\delta m^2$ (for small enough $\delta m^2$)
which means that the algorithm {\em eliminates
critical slowing down completely\/}. Moreover it was found that the number
of inverse iterations to calculate the smooth interpolation operators
was six, again independent of the problem parameters, so critical slowing down
does not enter through the back-door. A detailed discussion of this result can
be found in \cite{IJMPC}.

In applying the algorithm to fermions we used a block-factor of~3 instead
of the more usual 2 to preserve the symmetry of the staggered grid also
on the coarser layers \cite{Mack}. The grid length therefore should be
chosen as $L=2 \cdot 3^N$, where $N$ is the number of layers.
To improve convergence, we introduced an additional layer $\Lambda^{\rm add}$,
consisting of only four points, one for each pseudoflavour. This allows
for more interpolation operators that cover a large part of the grid.
(In the case of bosons, one could also use a block factor of three. There, the
additional layer is not needed, because we found that even for this block
factor $\tau$ was quite small and there was no critical slowing down.)

\begin{table}
\caption{Convergence rate $\tau$ of the ISU-algorithm applied to the squared
Dirac equation
as a function of the grid length for
physically scaled gauge fields.\label{TauFerm}
}

\begin{center}
\begin{tabular}{llll}
\hline
\noalign{\smallskip}
Grid size&$18^2$&$54^2$&$162^2$\\
\hline
\# runs&50&50&9\\
$\beta$&1&9&81\\
$\delta m^2$&$0.002$&$0.002/9$&$0.002/81$\\
$\tau$&$25.6\pm0.9$&$6.67\pm0.14 $&$7.5\pm0.8 $\\
\hline
\end{tabular}
\end{center}
\end{table}

Table~\ref{TauFerm} shows preliminary results for
the convergence rate for physically scaled
gauge fields, i.e.~for fields where $\beta\propto L^2$ and $\delta
m^2\propto L^{-2}$. The calculations were done on grids with grid size
$18^2$--$162^2$ with $\beta=1.0$ and $\delta m^2=2\cdot10^{-3}$ on the
smallest lattice. Obviously there is no critical slowing down in this
sense, which should be expected as the squared Dirac operator
approaches the Laplacian in the continuum limit.

However, for small $\beta$ the convergence rates are bad (large absolute
values) even on small lattices. The situation is much worse than for bosons,
where $\tau$ was practically independent of all parameters.

\section{Conclusions}
The great success of the algorithm in case of the Laplace equation
and the results obtained for the Dirac case pose three questions:

\begin{enumerate}
\item What is the crucial difference between the Laplacian and the Dirac
operator, causing the one to converge much better than the other?
\item Can the algorithm be improved, perhaps by including insights
  gained from the answer to the first question?
\item How does the algorithm (or its improved version) behave in four
dimensions?
\end{enumerate}

At the moment we are studying mainly the first question. It seems that
at least part of the answer lies in the shape of the low-lying eigenmodes
of the Dirac-operator.

\section*{Acknowledgements}
I wish to thank Gerhard Mack for his constant support. Further thanks
are due to A.  Brandt, S.  Solomon and all members of the Hamburg
Multigrid Group for helpfull discussions.

%
%
%

\end{document}